\documentclass[12pt]{iopart}
\usepackage{iopams}
\newcommand{\sh}[1]{e^{{#1}\partial_s}}

\newcommand{\defeq}{\stackrel{\mathrm{def}}{=}}
\newcommand{\eqref}[1]{(\ref{#1})}
\newcommand{\qed}{\hfill $\square$\smallskip}

\newcommand{\IC}{\mathbb{C}}
\newcommand{\IZ}{\mathbb{Z}}

\newtheorem{theorem}{Theorem}
\newtheorem{proposition}{Proposition}
\newtheorem{lemma}{Lemma}
\newtheorem{corollary}{Corollary}

\begin{document}
\title{
A $q$-analogue of $\widehat{\mathfrak{gl}}_3$ 
hierarchy and $q$-Painlev\'e VI}
\author{Saburo Kakei$^1$ and 
Tetsuya Kikuchi$^2$\footnote{
Present Address: 
Department of Mathematical Sciences,
University of Tokyo, Komaba, Meguro-ku, 
Tokyo 153, Japan.}}
\address{$^1$ Department of Mathematics, Rikkyo University, 
Nishi-ikebukuro, Toshima-ku, Tokyo 171-8501, Japan}
\ead{kakei@rkmath.rikkyo.ac.jp}
\bigskip

\address{$^2$ Mathematical Institute, Tohoku University,
Aoba-ku, Sendai 980-8578, Japan}
\ead{tkikuchi@math.tohoku.ac.jp}

\begin{abstract}
A $q$-analogue of the $\widehat{\mathfrak{gl}}_3$ 
Drinfel'd-Sokolov hierarchy is proposed as 
a reduction of the $q$-KP hierarchy.
Applying a similarity reduction 
and a $q$-Laplace transformation to the hierarchy, 
one can obtain the $q$-Painlev\'e VI equation 
proposed by Jimbo and Sakai. 
\end{abstract}

\submitto{\JPA}

\ \\

\textit{
Dedicated to Professors Junkichi Satsuma and 
Basil Grammaticos 
on the occasion of the 60th birthday.}

\maketitle

\section{Introduction}
In the preceding work \cite{KakeiKikuchi05}, we established 
a relationship between the generic Painlev\'e VI equation and 
the $\widehat{\mathfrak{gl}}_3$-Drinfel'd-Sokolov hierarchy
that contains the three-wave resonant system.
Our approach is based on a similarity reduction of the 
generalized Drinfel'd-Sokolov hierarchy that has been 
discussed in \cite{KakeiKikuchi04}.
We remark that 
Conte, Grundland and Musette also discussed a reduction from 
the three-wave resonant system to the generic 
Painlev\'e VI \cite{ConteGrundlandMusette}.

On the other hand, $q$-difference soliton equations
have been discussed by several researchers 
\cite{KS91,WZZ94,
Frenkel96,AHvM98,Iliev98,KNY02,Takasaki05}.
In \cite{KNY02}, Kajiwara, Noumi and Yamada discussed
a $q$-analogue of a similarity reduction from 
the $q$-KP hierarchy to $q$-Painlev\'e equations. 
The main purpose of the present article is to
obtain a $q$-analogue of the Painlev\'e VI equation 
as a similarity reduction of the multi-component $q$-KP hierarchy.

In \cite{JimboSakai96}, Jimbo and Sakai 
proposed a $q$-difference analogue of the sixth 
Painlev\'e equation ($q$-P$_\mathrm{VI}$), 
which is a coupled system of $q$-difference equations:
\begin{equation}
\label{qP6}
\left\{\begin{array}{l}
\displaystyle
\frac{y(t)y(qt)}{a_3a_4}=
\frac{\left\{z(qt)-tb_1\right\}\left\{z(qt)-tb_2\right\}}
{\left\{z(qt)-b_3\right\}\left\{z(qt)-b_4\right\}},\\
\displaystyle
\frac{z(t)z(qt)}{b_3b_4}=
\frac{\left\{y(t)-ta_1\right\}\left\{y(t)-ta_2\right\}}
{\left\{y(t)-a_3\right\}\left\{y(t)-a_4\right\}},
\end{array}\right.
\end{equation}
where the parameters $a_j$, $b_j$ ($j=1,2,3,4$) obey 
the constraint
\begin{equation}
\frac{b_1b_2}{b_3b_4}=q\frac{a_1a_2}{a_3a_4}.
\end{equation}
These equations are obtained from 
a connection preserving deformation of 
a linear $q$-difference equation, 
\begin{eqnarray}
Y(q\zeta,t)&=&
\mathcal{A}(\zeta;t)Y(\zeta,t), \quad
\mathcal{A}(\zeta;t)\defeq
\mathcal{A}_0(t)+\mathcal{A}_1(t)\zeta+\mathcal{A}_2(t)\zeta^2,
\label{Eq_Y(qx)}\\
Y(\zeta,qt)&=&
\frac{\zeta(\zeta I+\mathcal{B}_0(t))}{(\zeta-qta_1)(\zeta-qta_2)}
Y(\zeta,t),
\label{Eq_Y(qt)}
\end{eqnarray}
where $Y(\zeta,t)$ is a $2\times 2$-matrix valued function
with respect to $\zeta$ and $t$. 
The coefficient matrices $\mathcal{A}_j(t)$ ($j=0,1,2$) 
are assumed to satisfy the conditions, 
\begin{equation}
\begin{array}{l}
\mathcal{A}_2(t)=
\left[\begin{array}{cc}
\kappa_1&0\\ 0&\kappa_2
\end{array}\right],
\quad\mbox{Eigenvalues of }\mathcal{A}_0(t)
\mbox{ are }t\theta_1, t\theta_2,\\
\det\mathcal{A}(\zeta,t)=
\kappa_1\kappa_2(\zeta-ta_1)(\zeta-ta_2)(\zeta-a_3)(\zeta-a_4),
\end{array}
\end{equation}
where the parameters $\kappa_i$, $\theta_i$ ($i=1,2$) 
are given by
\begin{equation}
\kappa_1 =\frac{1}{qb_3},\quad
\kappa_2 =\frac{1}{b_4},\quad
\theta_1 =\frac{a_1a_2}{b_1},\quad
\theta_2 =\frac{a_1a_2}{b_2}.
\end{equation}

The variables $y(t)$, $z(t)$ of the $q$-P$_\mathrm{VI}$ are
related to the coefficient matrix 
$\mathcal{A}(\zeta;t)$ as follows:
\begin{eqnarray}
&&\left(\mathcal{A}(\zeta=y(t);t)\right)_{12}=0,\\
&&z(t) = 
\frac{(y-ta_1)(y-ta_2)}{q\left(\mathcal{A}
(\zeta=y;t)\right)_{11}}=
\frac{\left(\mathcal{A}
(\zeta=y;t)\right)_{22}}{q\kappa_1\kappa_2(y-a_3)(y-a_4)},
\end{eqnarray}
where $(M)_{ij}$ denotes the $(i,j)$-component of 
a matrix $M$.

In the next section, we introduce a $q$-analogue of 
the $\widehat{\mathfrak{gl}}_N$ hierarchy as a reduced case of the 
multi-component $q$-KP hierarchy based on the work \cite{Takasaki05}. 
We will show that the $q$-Painlev\'e VI can be obtained 
as a similarity reduction of 
the $q$-$\widehat{\mathfrak{gl}}_3$ hierarchy.

\section{A $q$-analogue of $\widehat{\mathfrak{gl}}_3$ hierarchy}
Throughout the paper, 
we assume $|q|>1$ unless mentioned otherwise.
We will use the following notations:
\begin{itemize}
 \item[](shift operator) \ $\sh{m}f(s)=f(s+m)$,
 \item[]($q$-shift operator) \ 
 $(T_q(z))^mf(z)=f(q^mz)$,
 \item[]($q$-difference operator) \ 
 $\displaystyle\mathcal{D}_q(z)f(z)
  =\frac{1-T_q(z)}{z}f(z)=\frac{f(z)-f(qz)}{z}$,
 \item[] ($q$-shifted factorial) \ 
 $\displaystyle (z;q^{-1})_n
  = \prod_{j=0}^{n-1}(1-q^{-j}z)$, \ 
 $\displaystyle (z;q^{-1})_\infty
  = \prod_{j=0}^\infty(1-q^{-j}z)$.
\end{itemize}

To describe a $q$-analogue of the multi-component 
KP hierarchy \cite{UenoTakasaki}, 
we define the Sato-Wilson operators, 
\begin{equation}
\label{Sato-Wilson_ops}
\eqalign{
W(\sh{};s,\underline{x})=
I+W_1\sh{-}+W_2\sh{-2}+\cdots,\cr
\bar{W}(\sh{};s,\underline{x})=
\bar{W}_0+\bar{W}_1\sh{}
+\bar{W}_2\sh{2}+\cdots.
}
\end{equation}
The coefficients $W_i=W_i(s;\underline{x})$ ($i=1,2,\ldots$), 
$\bar{W}_j=\bar{W}_j(s;\underline{x})$ ($j=0,1,2,\ldots$) are 
$N\times N$-matrix-valued functions that depend on 
a discrete variable $s$ and a set of parameters 
$\underline{x}=\{x_1^{(k)},x_2^{(k)},\ldots\,
(k=1,\dots,N)\}$.
We assume that $\bar{W}_0$ is invertible.

For a difference operator $A(\sh{})=\sum_n A_n\sh{n}$, 
we denote by $[A(\sh{})]_{\ge 0}$ the projection to 
the non-negative part:
$
 [A(\sh{})]_{\ge 0}=\sum_{n\ge 0}A_n\sh{n}.
$
We define a $q$-analogue of the Sato equation as 
\begin{equation}
\fl
\mathcal{D}_q(x_n^{(k)})\widetilde{W}
=\left[\Big(T_q(x_n^{(k)})W\Big)
 I_k\sh{n}W^{-1}\right]_{\ge 0}\widetilde{W}
-\Big(T_q(x_n^{(k)})\widetilde{W}\Big)I_k\sh{n}, 
\label{SatoEq}
\end{equation}
where $\widetilde{W}=W$ or $\bar{W}$, and 
$I_k=[\delta_{ij}\delta_{ik}]_{1\le i,j\le N}$.
We remark the hierarchy defined above is 
slightly different from that of \cite{Takasaki05}.
\begin{proposition}[Scaling symmetry]
\label{prop:ScalingSymmetry}
For a constant $\lambda\in\IC^\times$, we
define $W_\lambda$ and $\bar{W}_\lambda$ as
\begin{eqnarray}
W_\lambda(\sh{};s,\underline{x}) & \defeq &
\lambda^{s+D(\alpha)}\circ W(\sh{};s,\underline{x}_\lambda)
\circ\lambda^{-s-D(\alpha)},
\label{def_W_lambda}\\
\bar{W}_\lambda(\sh{};s,\underline{x}) & \defeq &
\lambda^{s+D(\alpha)}\circ\bar{W}(\sh{};s,\underline{x}_\lambda)
\circ\lambda^{-s-D(\beta)}, 
\end{eqnarray}
where 
$D(\alpha)=\mathrm{diag}[\alpha_1,\ldots,\alpha_N]$, 
$D(\beta)=\mathrm{diag}[\beta_1,\ldots,\beta_N]$ 
are constant matrices and $\underline{x}_\lambda=
\{\lambda x_1^{(k)},\lambda^2 x_2^{(k)},\ldots \, (k=1,\dots,N)\}$.
If $W$ and $\bar{W}$ solve the $q$-Sato equation 
\eqref{SatoEq}, so do $W_\lambda$ and $\bar{W}_\lambda$.
\end{proposition}
Proposition \ref{prop:ScalingSymmetry} can be checked by
a direct calculation.

We define formal Baker-Akhiezer functions, 
\begin{eqnarray}
\fl
\Psi_q^{(\infty)}(z;s,\underline{x})
&=& W(z;s,\underline{x})
\Psi_{q,0}^{(\infty)}(z;s,\underline{x}),\\
\fl
\Psi^{(\infty)}_{q,0}(z;s,\underline{x})
&=& z^{s+D(\alpha)}\prod_{j\ge 1}\mathrm{diag}\!\left[
(z^jx_j^{(1)}q^{-1};q^{-1})_\infty,\dots,
(z^jx_j^{(N)}q^{-1};q^{-1})_\infty\right],\\
\fl
\Psi_q^{(0)}(z;s,\underline{x})
&=& \bar{W}(z;s,\underline{x})
\Psi_{q,0}^{(0)}(z;s,\underline{x}),\\
\fl
\Psi^{(0)}_{q,0}(z;s,\underline{x})
&=& z^{s+D(\beta)}\prod_{j\ge 1}\mathrm{diag}\!\left[
(z^jx_j^{(1)}q^{-1};q^{-1})_\infty,\dots,
(z^jx_j^{(N)}q^{-1};q^{-1})_\infty\right],
\end{eqnarray}
where we have assumed that $|q|>1$ for convergence.
{}From \eqref{SatoEq}, 
it follows that both $\Psi^{(\infty)}_q(z;s,x)$ and 
$\Psi^{(0)}_q(z;s,x)$ satisfy the same 
$q$-difference equation of the form, 
\begin{equation}
\label{eq_for_BA}
\fl
\mathcal{D}_q(x_n^{(k)})
\Psi_q(z;s,\underline{x})
=\left[\Big(T_q(x_n^{(k)})W\Big)
I_k\sh{n}W^{-1}\right]_{\ge 0}
\Psi_q(z;s,\underline{x}).
\end{equation}

We now impose the condition,
\begin{equation}
\label{(1,1,1)-reduction}
\eqalign{
W(\sh{};s+1,\underline{x})
=W(\sh{};s,\underline{x}),
\cr
\bar{W}(\sh{};s+1,\underline{x})
=\bar{W}(\sh{};s,\underline{x}).
}
\end{equation}
If a difference operator $A(\sh{};s)$ satisfies the condition 
$A(\sh{};s+1)=A(\sh{};s)$, the correspondence 
\begin{equation}
\label{corresp}
A(\sh{};s)=\sum_{n\in\IZ}A_n(s)\sh{n}
 \quad\leftrightarrow\quad 
A(z;s)=\sum_{n\in\IZ}A_n(s) z^n
\end{equation}
preserves sums, products and commutators \cite{UenoTakasaki}. 
Here $z$ is used as a formal indeterminate. 
The $q$-Sato equation \eqref{SatoEq} 
then takes the following form:
\begin{eqnarray}
\mathcal{D}_q(x_n^{(k)})\widetilde{W} &=
C_n^{(k)}\widetilde{W}
-z^n(T_q(x_n^{(k)})\widetilde{W})I_k,
\quad \widetilde{W}=W,\bar{W}, 
\label{gl3-SatoEq}\\
C_n^{(k)}(z;\underline{x})&=
\left[z^n(T_q(x_n^{(k)})W)
I_kW^{-1}\right]_{\ge 0}.
\label{expression_Ck}
\end{eqnarray}
If we replace $x_n^{(k)}$ by $(1-q)x_n^{(k)}$ and 
take the limit $q\to 1$, 
the $q$-Sato equation \eqref{gl3-SatoEq} is reduced to 
the $\widehat{\mathfrak{gl}}_N$ hierarchy 
discussed in \cite{KakeiKikuchi05}. 
In this sense, we call as ``$q$-$\widehat{\mathfrak{gl}}_N$ hierarchy''
the hierarchy described by \eqref{gl3-SatoEq}.

Hereafter we restrict ourselves to the case $N=3$, 
and set $x_n^{(k)}=0$ for $n\ge 2$.
We will use the abbreviation $x_k=x_1^{(k)}$, 
$T_k=T_q(x_k)$, $C_k=C_1^{(k)}$ ($k=1,2,3$).
Then we can rewrite the $q$-Sato equation \eqref{gl3-SatoEq} 
as 
\begin{equation}
\label{gl3-SatoEq_2}
\left\{
-zx_kI_k+ V_k(\underline{x})
\right\}\widetilde{W}
=(T_k\widetilde{W})\left(-zx_kI_k+I\right),
\end{equation}
where $\widetilde{W}=W$ or $\bar{W}$, and 
$V_k(\underline{x})$ is defined by 
\begin{equation}
\label{def:V_k}
V_k(\underline{x})=
I-x_k\left\{(T_kW_1(\underline{x}))
I_k-I_kW_1(\underline{x})\right\}. 
\end{equation}
The matrix $V_k(\underline{x})$ is 
related to $C_k(z;\underline{x})$ as 
\begin{equation}
I-x_kC_k(z;\underline{x})
=-zx_kI_k+ V_k(\underline{x}).
\end{equation}
The concrete expressions of $V_k(z;\underline{x})$ ($k=1,2,3$) 
are given as follows:
\begin{eqnarray}
V_1(z;\underline{x})&=& I-x_1
\left[\begin{array}{ccc}
 T_1(w_{11}) - w_{11}  &  -w_{12} & -w_{13} \\
 T_1(w_{21}) &     0   &    0    \\
 T_1(w_{31}) &     0   &    0 
\end{array}\right],\label{C1}\\
V_2(z;\underline{x})&=& I-x_2
\left[\begin{array}{ccc}
     0   &  T_2(w_{12})  &    0 \\
 -w_{21} & T_2(w_{22})-w_{22}     & -w_{23} \\
     0   &  T_2(w_{32})  &    0 
\end{array}\right],\label{C2}\\
V_3(z;\underline{x})&=& I-x_3
\left[\begin{array}{ccc}
     0   &    0     & T_3(w_{13}) \\
     0   &    0     & T_3(w_{23}) \\
 -w_{31} & -w_{32}  & T_3(w_{33}) - w_{33}
\end{array}\right],\label{C3}
\end{eqnarray}
where $w_{ij}=w_{ij}(\underline{x})$ denotes the $(i,j)$-element 
of $W_1$.

The $q$-$\widehat{\mathfrak{gl}}_3$ hierarchy contains 
a $q$-analogue of the three-wave resonant system. 
To see this, we consider the reduced case of \eqref{eq_for_BA}, 
which does not depend on $s$:
\begin{equation}
\mathcal{D}_q(x_k)\Psi_q(z;\underline{x})
=C_k(z;\underline{x})
\Psi_q(z;\underline{x}).
\end{equation}
This can be rewritten as 
\begin{equation}
\label{eq_for_BA_gl3}
T_k\Psi_q(z;\underline{x})
=\left\{-zx_kI_k+ V_k(\underline{x})\right\}
\Psi_q(z;\underline{x}).
\end{equation}
{}From the compatibility condition of \eqref{eq_for_BA_gl3}, 
we have
\begin{equation}
\label{compatibility}
\left\{\begin{array}{l}
x_kI_kV_l + x_l(T_lV_k)I_l
= x_lI_lV_k + x_k(T_kV_l)I_k,\\
(T_kV_l)V_k = (T_lV_k)V_l,
\end{array}\right.
\end{equation}
for $k,l=1,2,3$.
Substituting \eqref{C1}, \eqref{C2}, \eqref{C3} for 
\eqref{compatibility}, one obtains
\begin{equation}
\label{q-3wave}
\mathcal{D}_q(x_k)w_{ij}=(T_kw_{ik})w_{kj}.
\end{equation}
If we impose the condition $w_{ji}=w_{ij}^*$ 
(complex conjugate of $w_{ij}$), 
the equations \eqref{q-3wave} can be regarded as 
a $q$-analogue of the three-wave resonant system.

In what follows, 
the matrix $\bar{W}_0(\underline{x})$ plays a 
crucial role. We prepare several lemmas for latter use.
\begin{lemma}
\label{lemma:det_w0}
Under the reduction condition \eqref{(1,1,1)-reduction}, 
$\bar{W}_0(\underline{x})$ satisfies
$
T_k \det\bar{W}_0(\underline{x})
 = \det\bar{W}_0(\underline{x})
$.
\end{lemma}
\textit{Proof.} 
The $\widehat{\mathfrak{gl}}_3$ $q$-Sato equation \eqref{gl3-SatoEq}
implies 
\begin{equation}
\fl\{T_aW(z)\}(I-zx_aI_a)\{W(z)\}^{-1}
=\{T_a\bar{W}(z)\}(I-zx_aI_a)\{\bar{W}(z)\}^{-1}.
\end{equation}
It follows that 
\begin{equation}
\det[T_aW(z)]\cdot
\{\det W(z)\}^{-1} =
\det[T_a\bar{W}(z)]\cdot
\{\det\bar{W}(z)\}^{-1}.
\end{equation}
There are no positive powers with respect to $z$
on the left-hand side, while no negative powers 
on the right-hand side.
Thus we obtain the result from the degree $0$ term.
\qed

\begin{lemma}
\label{lemma:T_kw0}
The matrix $\bar{W}_0(\underline{x})$ satisfies 
$
T_k\bar{W}_0(\underline{x})
=V_k(\underline{x})\bar{W}_0(\underline{x})
$
\end{lemma}
\textit{Proof.} 
This is a direct consequence of 
\eqref{gl3-SatoEq_2} with $\widetilde{W}=\bar{W}$.
\qed

\begin{lemma}
\label{lemma:det[V+lambda_I]}
$\det\left[V_k+\lambda I_k\right]=1+\lambda$ ($k=1,2,3$).
\end{lemma}
\textit{Proof.} 
The case with $\lambda=0$ follows from Lemmas 1 and 2.
Using this result, we have
\begin{equation*}
\det\left[V_k+\lambda I_k\right]
=\det\left[V_k\right]+\lambda\times
\left\{\mbox{the }(k,k)\mbox{-cofactor of }V_k\right\}
=1+\lambda,
\end{equation*}
where we have used 
\eqref{C1}, \eqref{C2} and \eqref{C3}.
\qed

\section{Similarity reduction to $q$-Painlev\'e VI}
\subsection{Similarity reduction to $q$-Schlesinger system}
Motivated by the scaling symmetry 
(Proposition \ref{prop:ScalingSymmetry}), we impose the following 
conditions on $W$ and $\bar{W}$, which we call 
``similarity conditions'':
\begin{eqnarray}
W(\sh{};s,\underline{x}) & = &
q^{s+D(\alpha)}\circ W(\sh{};s,\underline{x}_q)
\circ q^{-s-D(\alpha)},
\label{similarity_W}\\
\bar{W}(\sh{};s,\underline{x}) & = &
q^{s+D(\alpha)}\circ\bar{W}(\sh{};s,\underline{x}_q)
\circ q^{-s-D(\beta)}.
\label{similarity_Wbar}
\end{eqnarray}
Under the reduction condition \eqref{(1,1,1)-reduction}, 
the similarity conditions 
\eqref{similarity_W}, \eqref{similarity_Wbar}
take the form 
\begin{eqnarray}
W(z;\underline{x}) & = &
q^{D(\alpha)}W(q^{-1}z;\underline{x}_q)q^{-D(\alpha)},
\label{gl3-similarity_W}\\
\bar{W}(z;\underline{x}) & = &
q^{D(\alpha)}\bar{W}(q^{-1}z;\underline{x}_q)q^{-D(\beta)}.
\label{gl3-similarity_Wbar}
\end{eqnarray}
The similarity condition for 
for $\bar{W}_0(\underline{x})$ follows from 
\eqref{gl3-similarity_Wbar}:
\begin{equation}
\label{similarity_w0}
\bar{W}_0(\underline{x}) =
q^{D(\alpha)}\bar{W}_0(\underline{x}_q)q^{-D(\beta)}.
\end{equation}
We remark that 
the parameters $\alpha_i$, $\beta_i$ ($i=1,2,3$)
should obey the relation, 
\begin{equation}
\alpha_1+\alpha_2+\alpha_3=\beta_1+\beta_2+\beta_3, 
\end{equation}
due to Lemma \ref{lemma:det_w0}.

The similarity conditions \eqref{gl3-similarity_W}, 
\eqref{gl3-similarity_Wbar} imply the following relation for 
$\Psi_q(z,\underline{x})$:
\begin{equation}
\Psi_q(qz,\underline{x})
=q^{D(\alpha)}\Psi_q(z,\underline{x}_q).
\end{equation}
Applying \eqref{eq_for_BA_gl3}, 
we can calculate $\Psi_q(z,\underline{x}_q)$ as
\begin{equation}
\fl
\Psi_q(z,\underline{x}_q) =
\left\{-zx_1I_1+ (T_2T_3V_1)\right\}
\left\{-zx_2I_2+ (T_3V_2)\right\}
\left\{-zx_3I_3+ V_3\right\}
\Psi_q(z,\underline{x}).
\end{equation}
Due to \eqref{gl3-SatoEq_2}, one can rewrite 
this equation in two different ways:
\begin{eqnarray}
\lefteqn{
\left\{-zx_1I_1+ (T_2T_3V_1)\right\}
\left\{-zx_2I_2+ (T_3V_2)\right\}
\left\{-zx_3I_3+ V_3\right\}
}\qquad&
\nonumber\\
&=&(T_1T_2T_3W)
\prod_{k=1}^3\left(I-zx_kI_k\right)
W^{-1}
\nonumber\\
&=&(T_1T_2T_3\bar{W})
\prod_{k=1}^3\left(I-zx_kI_k\right)
\bar{W}^{-1}.
\label{expression2_for_prod(1-xC)}
\end{eqnarray}
Since \eqref{expression2_for_prod(1-xC)} has no negative 
powers with respect to $z$, we have
\begin{eqnarray}
\lefteqn{(T_1T_2T_3W)
\prod_{k=1}^3\left(I-zx_kI_k\right)
W^{-1}}\quad\nonumber\\
&=\left[(T_1T_2T_3W)
\prod_{k=1}^3\left(I-zx_kI_k\right)
W^{-1}\right]_{\ge 0}
\nonumber\\
&= \left[
(T_1T_2T_3W)\left(
I-z\sum_{k=1}^3x_kI_k\right)
W^{-1}\right]_{\ge 0}
\nonumber\\
&=I-\sum_{k=1}^3x_k
\left[z(T_1T_2T_3W)I_kW^{-1}\right]_{\ge 0}.
\end{eqnarray}
It follows that \eqref{expression2_for_prod(1-xC)} has
the following expression:
\begin{equation}
\label{T1T2T3W}
(T_1T_2T_3\bar{W})
\prod_{k=1}^3\left(I-zx_kI_k\right)
\bar{W}^{-1}
=U(\underline{x})-z\sum_{k=1}^3x_kI_k,
\end{equation}
where $U(\underline{x})$ is a $3\times 3$ matrix.
Comparing the $z^0$-terms in \eqref{T1T2T3W}
and using \eqref{similarity_w0}, we get
\begin{equation}
U(\underline{x})=(T_1T_2T_3\bar{W}_0)\bar{W}_0^{-1}
=q^{-D(\alpha)}
\bar{W}_0q^{D(\beta)}
\bar{W}_0^{-1}.
\end{equation}
Thus we have obtained a linear $q$-difference 
equation for $\Psi_q$:
\begin{equation}
\label{3by3_linear_1}
\Psi_q(qz;\underline{x})
= \left\{-z\,q^{D(\alpha)}X
+ \bar{W}_0q^{D(\beta)}\bar{W}_0^{-1}\right\}
\Psi_q(z;\underline{x}),
\end{equation}
where $X=\mathrm{diag}[x_1,x_2,x_3]$. 
It is convenient to introduce a gauge-transformed function
$\widetilde{\Psi}_q\defeq \bar{W}_0^{-1}\Psi_q$ 
that satisfies the following system of equations:
\begin{eqnarray}
\widetilde{\Psi}_q(qz;\underline{x})
&=& \left\{-z\bar{W}_0^{-1}q^{D(\alpha)}
X\bar{W}_0+ q^{D(\beta)}\right\}
\widetilde{\Psi}_q(z;\underline{x}),
\label{qSystem_3times3_linear}\\
T_k\widetilde{\Psi}_q(z;\underline{x})
&=& \left\{-zx_k
(T_k\bar{W}_0)^{-1}I_k\bar{W}_0+ I\right\}
\widetilde{\Psi}_q(z;\underline{x}).
\label{qSystem_3times3_deformation}
\end{eqnarray}
As we shall show in what follows, 
the system of the linear $q$-difference equations 
\eqref{qSystem_3times3_linear}, 
\eqref{qSystem_3times3_deformation} 
works as a Lax pair for the $q$-P$_\mathrm{VI}$ 
with $3\times 3$-matrix coefficients, 
which is a $q$-analogue of the formulation 
used in \cite{Har,Maz,Boalch05}.

To establish a link between the $3\times 3$-matrix system 
\eqref{qSystem_3times3_linear}, 
\eqref{qSystem_3times3_deformation} and 
the $2\times 2$-matrix system \eqref{Eq_Y(qx)}, \eqref{Eq_Y(qt)}, 
we use a $q$-analogue of Laplace transform due to Hahn \cite{Hahn}.
For a function $f(z)$, we define 
$\mathcal{L}_q[f](\zeta)$ and $\mathcal{L}_q^{-1}[f](z)$ as
\begin{eqnarray}
\mathcal{L}_q[f](\zeta)
 &= \frac{1}{\zeta}
    \sum_{n=0}^\infty
     \frac{q^{-n} f(\zeta^{-1}q^{-n})}{(q^{-1};q^{-1})_n},
 \label{q-Laplace}\\
\mathcal{L}_q^{-1}[f](z)
 &= \frac{1}{z}
    \sum_{n=0}^\infty
    \frac{(-1)^n q^{-n(n-1)/2}f(z^{-1}q^{n})}{(q^{-1};q^{-1})_n}.
 \label{Inv_q-Laplace}
\end{eqnarray}
The transformations \eqref{q-Laplace}, \eqref{Inv_q-Laplace} 
have the following properties:
\begin{eqnarray}
\mathcal{L}_q[\mathcal{D}_{q^{-1}}(z)f(z)](\zeta)
 = \zeta\mathcal{L}_q[f(z)](\zeta)-(q^{-1};q^{-1})_\infty^{-1}f(0),
\label{formula1}\\
\mathcal{L}_q[zf(z)](\zeta)
 = \mathcal{D}_{q}(\zeta)\mathcal{L}_q[f(z)](\zeta),
\label{formula2}\\
\mathcal{L}^{-1}_q[\mathcal{D}_{q}(\zeta)f(\zeta)](z)
 = z\mathcal{L}^{-1}_q[f(\zeta)](z),
\label{formula3}\\
\mathcal{L}^{-1}_q[\zeta f(\zeta)](z)
 = \mathcal{D}_{q^{-1}}(z)\mathcal{L}^{-1}_q[f(\zeta)](z),
\label{formula4}\\
\mathcal{L}^{-1}_q[\mathcal{L}_q[f]](z)=f(z),
\label{inversion1}\\
\mathcal{L}_q[\mathcal{L}^{-1}_q[f]](\zeta)=f(\zeta).
\label{inversion2}
\end{eqnarray}
We outline a proof of \eqref{formula1}--\eqref{inversion2} 
in Appendix.

If we define $\widetilde{\Phi}_q(z)
=\mathcal{L}^{-1}_q[\widetilde{\Psi}_q(\zeta)](z)$, 
we can show that the transformed function 
$\widetilde{\Phi}_q(\zeta)$ satisfies the linear equations,
\begin{eqnarray}
\mathcal{D}_{q^{-1}}(\zeta)
\widetilde{\Phi}_q(\zeta;\underline{x})
&=& \sum_{j=1}^3\frac{\bar{W}_0^{-1}I_j\bar{W}_0
(I-q^{D(\beta)+I})}{\zeta-q^{\alpha_j+1}x_j}
\widetilde{\Phi}_q(\zeta;\underline{x}),
\label{Eq_tildePhi}\\
\mathcal{D}_q(x_k)\widetilde{\Phi}_q(\zeta;\underline{x})
&=& \frac{(T_k\bar{W}_0)^{-1}I_k
\bar{W}_0(I-q^{D(\beta)+I})
}{\zeta-q^{\alpha_k+1}x_k}
\widetilde{\Phi}_q(\zeta;\underline{x}).
\label{before_projection_deformEq}
\end{eqnarray}
We can set $\beta_3=-1$ without loss of generality. 
With this choice, we have 
$(I-q^{D(\beta)+I})_{j3}= 0$ ($j=1,2,3$)
and we can restrict the equations 
\eqref{Eq_tildePhi}, \eqref{before_projection_deformEq} 
to the two dimensional subspace 
$\{\,^t(\widetilde{\phi}_1,\widetilde{\phi}_2,0)\,\}$.
Thus we obtain the $2\times 2$ system of the form,
\begin{eqnarray}
\mathcal{D}_{q^{-1}}(\zeta)
\widetilde{Y}(\zeta;\underline{x})
&=& -\sum_{j=1}^3\frac{A_j(\underline{x})
}{\zeta-q^{\alpha_j+1}x_j}
\widetilde{Y}(\zeta;\underline{x}),
\label{projection_Eq_tildePhi}\\
\mathcal{D}_q(x_k)
\widetilde{Y}(\zeta;\underline{x})
&=& -\frac{B_k(\underline{x})}{\zeta-q^{\alpha_k+1}x_k}
\widetilde{Y}(\zeta;\underline{x}),
\label{projection_deformEq}
\end{eqnarray}
where $A_k(\underline{x})$, 
$B_k(\underline{x})$ ($k=1,2,3$) are defined by
\begin{eqnarray}
\fl
A_k(\underline{x}) &=&
\left[\!\!\begin{array}{c}
(\bar{W}_0^{-1})_{1k}\\ (\bar{W}_0^{-1})_{2k}
\end{array}\!\!\right]
\Big[(\bar{W}_0)_{k1}\;
(\bar{W}_0)_{k2}\Big]
\left[\begin{array}{cc}
q^{\beta_1+1}-1&0\\ 0&q^{\beta_2+1}-1
\end{array}\right],
\label{def_Ak}\\
\fl
B_k(\underline{x}) &=&
\left[\!\!\begin{array}{c}
((T_k\bar{W}_0)^{-1})_{1k}\\
((T_k\bar{W}_0)^{-1})_{2k}
\end{array}\!\!\right]
\Big[(\bar{W}_0)_{k1}\;
(\bar{W}_0)_{k2}\Big]
\left[\begin{array}{cc}
q^{\beta_1+1}-1&0\\ 0&q^{\beta_2+1}-1
\end{array}\right].
\label{def_Bk}
\end{eqnarray}
We remark that the matrices $A_1,A_2,A_3$ satisfy the relation
\begin{equation}
\label{sum_Ak}
A_1+A_2+A_3+I = 
\left[\begin{array}{cc}
q^{\beta_1+1}&0\\ 0&q^{\beta_2+1}
\end{array}\right].
\end{equation}
We call \eqref{projection_Eq_tildePhi} and 
\eqref{projection_deformEq} as $q$-Schlesinger system 
since the limiting case $q\to 1$ coincides with 
the Schlesinger system associated with P$_\mathrm{VI}$.

\subsection{Relation with the $q$-Painlev\'e VI}
Hereafter we set $\beta_3=-1$, $x_3=0$. 
We introduce $Y(\zeta;\underline{x})$ as 
\begin{equation}
Y(\zeta;\underline{x})=\frac{
(q^{\alpha_1}x_1\zeta^{-1};q^{-1})_\infty
(q^{\alpha_2}x_2\zeta^{-1};q^{-1})_\infty}{
(\zeta;q^{-1})_\infty^2 
(q^{-1}\zeta^{-1};q^{-1})_\infty^2}
\widetilde{Y}(\zeta;\underline{x}).
\label{def_Ytilde}
\end{equation}
{}From \eqref{projection_Eq_tildePhi}, 
\eqref{projection_deformEq}, we have
\begin{eqnarray}
Y(q^{-1}\zeta;\underline{x})
&=& \mathcal{A}(\zeta;\underline{x})
Y(\zeta;\underline{x}),
\label{Eq_Y(zeta/q)}\\
T_kY(\zeta;\underline{x})
&=& \zeta^{-1}\left\{(\zeta-q^{\alpha_k+1}x_k)I
+x_kB_k(x)\right\}Y(\zeta;\underline{x}),
\label{deformation_JS_0}
\end{eqnarray}
where the coefficient matrix $\mathcal{A}(\zeta;\underline{x})$ 
is given by
\begin{eqnarray}
\mathcal{A}(\zeta;\underline{x}) 
&=& (\zeta -q^{\alpha_1+1}x_1)
     (\zeta -q^{\alpha_2+1}x_2)
     (I+A_3(x))\nonumber\\
& &
 + \zeta(\zeta -q^{\alpha_2+1}x_2)A_1(x)
 + \zeta(\zeta -q^{\alpha_1+1}x_1)A_2(x).
\label{def:calA(zeta;x)}
\end{eqnarray}

The coefficient matrix $\mathcal{A}(\zeta;\underline{x})$ 
has the form 
\begin{equation}
\mathcal{A}(\zeta;\underline{x})
=\mathcal{A}_2\zeta^2
 + \mathcal{A}_1\zeta + \mathcal{A}_0,
\end{equation}
where the matrices 
$\mathcal{A}_k=\mathcal{A}_k(\underline{x})$ ($k=0,1,2$) 
are given by
\begin{equation}
\eqalign{
\mathcal{A}_0 =
q^{\alpha_1+\alpha_2+2}x_1x_2(I+A_3),\quad
\mathcal{A}_2=\mathrm{diag}\!\left[
q^{\beta_1+1},\, q^{\beta_2+1}\right],\cr
\mathcal{A}_1 =
-(q^{\alpha_1+1}x_1+q^{\alpha_2+1}x_2)\mathcal{A}_2
+q^{\alpha_1+1}x_1A_1+q^{\alpha_2+1}x_2A_2
}
\label{def_calA}
\end{equation}

\begin{proposition}
Eigenvalues of $\mathcal{A}_0$ are 
$x_1x_2 q^{\alpha_1+\alpha_2+2}$, 
$x_1x_2 q^{\alpha_1+\alpha_2+\alpha_3+3}$.
\end{proposition}
\textit{Proof.} 
Denote as $F(\lambda)$ 
the characteristic polynomial of $\mathcal{A}_0$:
\begin{equation}
F(\lambda)=\det\left[\lambda I-\mathcal{A}_0\right]
=\det\left[\tilde{\lambda}I
-q^{\alpha_1+\alpha_2+2}x_1x_2A_3\right], 
\end{equation}
where we have set 
$\tilde{\lambda}=\lambda-q^{\alpha_1+\alpha_2+2}x_1x_2$.
Using the fact
\begin{equation}
\left(\bar{W}_0^{-1}I_3\bar{W}_0
(q^{D(\beta)+I}-I)\right)_{ij}
=\left\{\begin{array}{ll}
(A_3)_{ij} & (1\le i,j\le 2),\\
0 & (j=3),
\end{array}\right.
\end{equation}
we can rewrite $\tilde{\lambda}F(\tilde{\lambda})$ 
in terms of a $3\times 3$ determinant:
\begin{eqnarray}
\tilde{\lambda}F(\tilde{\lambda})
&=& \det\left[\tilde{\lambda}I
-q^{\alpha_1+\alpha_2+2}x_1x_2
\left\{\bar{W}_0^{-1}I_3\bar{W}_0
(q^{D(\beta)+I}-I)\right\}\right]
\nonumber\\
&=& \det\left[\tilde{\lambda}I
-q^{\alpha_1+\alpha_2+2}x_1x_2
I_3\left(\bar{W}_0
q^{D(\beta)+I}\bar{W}_0^{-1}-I\right)\right].
\end{eqnarray}
{}From \eqref{def:V_k} and 
\eqref{similarity_w0} with Lemma \ref{lemma:T_kw0}, 
it follows that 
\begin{eqnarray}
\lefteqn{I_3 
\bar{W}_0q^{D(\beta)+I}
\bar{W}_0^{-1}
=q^{D(\alpha)+I}I_3 (T_1T_2\bar{W}_0)
\bar{W}_0^{-1}
}\quad\nonumber\\
&=& q^{D(\alpha)+I}
I_3 (T_2V_1)V_2\nonumber\\
&=& q^{D(\alpha)+I}\{
I_3-x_1I_3(T_1T_2W_1)I_1
-x_2I_3(T_2W_1)I_2\nonumber\\
&&\qquad\qquad\qquad
 +x_1x_2I_3(T_1T_2W_1)I_1
(T_2W_1)I_2\}.
\end{eqnarray}
Thus we have $\left(I_3 
\bar{W}_0q^{D(\beta)+I}
\bar{W}_0^{-1}\right)_{33}=q^{\alpha_3+1}$ and obtain
\begin{equation}
\tilde{\lambda}F(\tilde{\lambda})
 =\tilde{\lambda}^2 \left\{\tilde{\lambda}-
  q^{\alpha_1+\alpha_2+2}x_1x_2
  (q^{\alpha_3+1}-1)\right\},
\end{equation}
which proves the proposition.
\qed

\begin{proposition}
$\det[\mathcal{A}(\zeta;\underline{x})]
= q^{\alpha_1+\alpha_2+\alpha_3+3}
\prod_{j=1}^2
(\zeta-x_j)(\zeta-q^{\alpha_j+1}x_j)$.
\end{proposition}
\textit{Proof.} 
Due to \eqref{def_Ak} and \eqref{def:calA(zeta;x)}, 
$\det[\mathcal{A}(\zeta;\underline{x})]$ can be written as
a $3\times 3$-determinant:
\begin{eqnarray}
\fl
\det\left[\mathcal{A}(\zeta;\underline{x})\right]
&=&(\zeta -q^{\alpha_1+1}x_1)^2
(\zeta -q^{\alpha_2+1}x_2)^2\nonumber\\
\fl
&& \times \det\Big[
I+
\bar{W}_0^{-1}I_3\bar{W}_0
(q^{D(\beta)+I}-I)\nonumber\\
\fl
&& \qquad\qquad
 + \zeta(\zeta -q^{\alpha_1+1}x_1)^{-1}
\bar{W}_0^{-1}I_1\bar{W}_0
(q^{D(\beta)+I}-I)\nonumber\\
\fl
&& \qquad\qquad
 + \zeta(\zeta -q^{\alpha_2+1}x_2)^{-1}
\bar{W}_0^{-1}I_2\bar{W}_0
(q^{D(\beta)+I}-I)\Big]
\nonumber\\
\fl
&=&(\zeta -q^{\alpha_1+1}x_1)^2
(\zeta -q^{\alpha_2+1}x_2)^2\nonumber\\
\fl
&& \times \det\Big[
I+
I_3\bar{W}_0
(q^{D(\beta)+I}-I)
\bar{W}_0^{-1}\nonumber\\
\fl
&& \qquad\qquad
 + \zeta(\zeta -q^{\alpha_1+1}x_1)^{-1}
I_1\bar{W}_0
(q^{D(\beta)+I}-I)
\bar{W}_0^{-1}\nonumber\\
\fl
&& \qquad\qquad
 + \zeta(\zeta -q^{\alpha_2+1}x_2)^{-1}
I_2\bar{W}_0
(q^{D(\beta)+I}-I)
\bar{W}_0^{-1}\Big].
\label{det[A_3times3]}
\end{eqnarray}
Applying the similarity condition \eqref{similarity_w0}
to \eqref{det[A_3times3]}, we have
\begin{eqnarray}
\det\left[\mathcal{A}(\zeta;\underline{x})\right]
&=&
q^{\alpha_1+\alpha_2+\alpha_3+3}\zeta^2
(\zeta -q^{\alpha_1+1}x_1)
(\zeta -q^{\alpha_2+1}x_2)
\nonumber\\
&&\quad \times
\det\left[
(T_2V_1)V_2
-\zeta^{-1}x_1I_1-\zeta^{-1}x_2I_2
\right].
\label{det[calA_(VV-I-I)]}
\end{eqnarray}
Furthermore, from \eqref{def:V_k}, it follows that
\begin{equation}
\fl (T_2V_1)V_2
-\zeta^{-1}x_1I_1-\zeta^{-1}x_2I_2
=\left(
T_2V_1-\zeta^{-1}x_1I_1
\right)
\left(
V_2-\zeta^{-1}x_2I_2
\right).
\label{farctorization_(VV-I-I)}
\end{equation}
According to 
\eqref{det[calA_(VV-I-I)]}, \eqref{farctorization_(VV-I-I)} 
and Lemma \ref{lemma:det[V+lambda_I]}, we obtain the result.
\qed

Next we consider the coefficient matrix of \eqref{deformation_JS_0}
with $k=1$.
\begin{lemma}
\label{lemma:det[I+B]}
$\det\left[
(\zeta-q^{\alpha_1+1}x_1)I
+x_1B_1(x)\right]=
(\zeta-x_1)(\zeta -q^{\alpha_1+1}x_1)$.
\end{lemma}
\textit{Proof.} 
Due to \eqref{def_Bk}, the determinant above can be written as
a $3\times 3$-determinant:
\begin{eqnarray}
\fl
\lefteqn{\det\left[
(\zeta-q^{\alpha_1+1}x_1)I
+x_1B_1(x)\right]}\quad\nonumber\\
\fl
&=&(\zeta-q^{\alpha_1+1}x_1)^{-1}\det\left[
(\zeta-q^{\alpha_1+1}x_1)I
+x_1(T_1\bar{W}_0)^{-1}I_1\bar{W}_0
 (q^{D(\beta)+I}-I)
\right]\nonumber\\
\fl
&=&(\zeta-q^{\alpha_1+1}x_1)^{-1}\det\left[
(\zeta-q^{\alpha_1+1}x_1)I
+x_1V_1^{-1}I_1
(\bar{W}_0q^{D(\beta)+I}\bar{W}_0^{-1}-I)
\right]\nonumber\\
\fl
&=&(\zeta-q^{\alpha_1+1}x_1)^{-1}\det\left[
(\zeta-q^{\alpha_1+1}x_1)I
+x_1I_1
\{q^{D(\alpha)+I}(T_1V_2)-V_1^{-1}\}
\right],
\end{eqnarray}
where we have used \eqref{similarity_w0} in the final line.
The result follows from a direct computation with 
\eqref{def:V_k}.
\qed

Now we are in position to state our main result.
\begin{theorem}
\label{thm:mainresult}
Assume that 
$W(z;\underline{x})$ and $\bar{W}(z;\underline{x})$ 
solve the $q$-Sato equation \eqref{gl3-SatoEq_2}, and satisfy
the similarity conditions \eqref{gl3-similarity_W},
\eqref{gl3-similarity_Wbar} with $\beta_3=-1$.
Take $Y^{(\infty)}_q(\zeta;x_1,x_2)$ as the $2\times 2$-matrix 
valued function associated with $W(z;\underline{x})$, 
and $Y^{(0)}_q(\zeta;x_1,x_2)$ with $\bar{W}(z;\underline{x})$.
If we replace $q$ by $q^{-1}$ and set $x_1=\gamma t$, 
then the functions 
\begin{equation}
Y^{(*)}(\zeta,t)
=\left[Y_q^{(*)}(\zeta;\gamma t,x_2)\right]_{q\to q^{-1}}
\quad (*=\infty, 0)
\end{equation}
solve the $q$-difference system
\eqref{Eq_Y(qx)}, \eqref{Eq_Y(qt)}.
The parameters are identified as follows:
\begin{equation}
\fl\eqalign{
\kappa_1 =q^{-\beta_1-1}, \quad \kappa_2 =q^{-\beta_2-1},
\quad \theta_1 = \gamma x_2 q^{-\alpha_1-\alpha_2-2}, \quad
\theta_2 = \gamma x_2 q^{-\alpha_1-\alpha_2-\alpha_3-3},\cr
a_1 = \gamma, \qquad a_2 = \gamma q^{-\alpha_1-1}, \qquad
a_3 = x_2, \qquad a_4 = x_2q^{-\alpha_2-1}.
}
\end{equation}
\end{theorem}
\textit{Proof.} 
We have already proved that both 
$Y_q^{(\infty)}(\zeta;\underline{x})$ and 
$Y_q^{(0)}(\zeta;\underline{x})$ solve
\eqref{Eq_Y(zeta/q)} with \eqref{def_calA}. 
The coefficient matrix $\mathcal{A}(\zeta;\underline{x})$ 
satisfies the desirous condition as shown in 
Propositions 2 and 3. 
The remaining task is to rewrite \eqref{deformation_JS_0} 
as \eqref{Eq_Y(qt)}. 
Using Lemma \ref{lemma:det[I+B]} to calculate the 
inverse of the coefficient matrix of \eqref{deformation_JS_0}, 
we get 
\begin{equation}
\tilde{Y}(\zeta;\underline{x})
=\frac{\zeta
\left\{(\zeta-x_1q^{\alpha_1+1})I
+x_1\widetilde{B}_1(\underline{x})\right\}
}{(\zeta-x_1)(\zeta -x_1q^{\alpha_1+1})}
 T_1\tilde{Y}(\zeta;\underline{x}),
\label{inverse_Y(qx_1)}
\end{equation}
where $\widetilde{B}_1(\underline{x})$ is 
defined by 
\begin{equation}
\widetilde{B}_1(\underline{x})
=\left[\!\!\begin{array}{cc}
(1-q^{\beta_2+1})(\bar{W}_0)_{12}\\
-(1-q^{\beta_1+1})(\bar{W}_0)_{11}
\end{array}\!\!\right]
\left[\!\!\begin{array}{cc}
(T_j\bar{W}_0^{-1})_{21} \;
-(T_j\bar{W}_0^{-1})_{11}
\end{array}\!\!\right].
\end{equation}
Applying $T_1^{-1}$ to \eqref{inverse_Y(qx_1)}, 
we obtain
\begin{equation}
T_1^{-1}\tilde{Y}(\zeta;x)
=\frac{\zeta
\left\{(\zeta-x_1q^{\alpha_1})I
+q^{-1}x_1(T_1^{-1}\widetilde{B}_1(\underline{x}))\right\}
}{(\zeta-q^{-1} x_1)(\zeta -x_1q^{\alpha_1})}
\tilde{Y}(\zeta;x).
\label{before_q->q^{-1}}
\end{equation}
If we replace $q$ by $q^{-1}$ and 
set $\mathcal{B}_0 = -x_1q^{-\alpha_1}I
+qx_1 (T_1[\widetilde{B}_1]_{q\to q^{-1}})$, then 
the equation \eqref{before_q->q^{-1}} agrees 
with \eqref{Eq_Y(qt)}. 
\qed.

\begin{corollary}
\label{cor:solutions}
Under the assumption of Theorem \ref{thm:mainresult}, 
we can obtain a solution of the $q$-P$_\mathrm{VI}$ 
written in terms of $\bar{W}_0$:
\begin{eqnarray}
\fl 
y &=&
\left[-\frac{(\mathcal{A}_0)_{12}}{(\mathcal{A}_1)_{12}}
\right]_{q\to q^{-1}},\\
\fl 
z &=& \left[
\frac{\{(\mathcal{A}_0)_{12} + x_1 (\mathcal{A}_1)_{12}\}
\{(\mathcal{A}_0)_{12} + x_1 q^{\alpha_1+1} (\mathcal{A}_1)_{12}\}}
{q(\mathcal{A}_1)_{12}
\left\{ (\mathcal{A}_0)_{11} (\mathcal{A}_1)_{12}
  - (\mathcal{A}_1)_{11} (\mathcal{A}_0)_{12}\right\}
  + q^{\beta_2+2} ((\mathcal{A}_0)_{12})^2 }
\right]_{q\to q^{-1}},
\end{eqnarray}
with
\begin{eqnarray}
\fl 
(\mathcal{A}_0)_{12}
&=& q^{\alpha_1 + \alpha_2+2}(q^{\beta_2+1}-1)x_1x_2 
(\bar{W}_0^{-1})_{13}(\bar{W}_0)_{32},
\\
\fl 
(\mathcal{A}_1)_{12}
&=& (q^{\beta_2+1}-1)
\left\{q^{\alpha_1 + 1}x_1 
    (\bar{W}_0^{-1})_{11}(\bar{W}_0)_{12}
  +q^{\alpha_2 + 1}x_2 
    (\bar{W}_0^{-1})_{12}(\bar{W}_0)_{22}\right\},
\\
\fl 
(\mathcal{A}_0)_{11}
&=& q^{\alpha_1 + \alpha_2+2}x_1x_2 
  \left\{1 + (q^{\beta_1+1}-1)
(\bar{W}_0^{-1})_{13}(\bar{W}_0)_{31}\right\},
\\
\fl 
(\mathcal{A}_1)_{11}
&=&
-q^{\beta_1+1}(q^{\alpha_1+1}x_1+q^{\alpha_2+1}x_2)
\nonumber\\
\fl 
&&
+(q^{\beta_1+1}-1)
\left\{
q^{\alpha_1+1}x_1
(\bar{W}_0^{-1})_{11}(\bar{W}_0)_{11}
+q^{\alpha_2+1}x_2
(\bar{W}_0^{-1})_{12}(\bar{W}_0)_{21}
\right\}.
\end{eqnarray}
\end{corollary}

\section{Concluding remarks}
In this paper, we have obtained the $q$-Painlev\'e VI \eqref{qP6}
as a similarity reduction of the $q$-$\widehat{\mathfrak{gl}}_3$ 
hierarchy \eqref{gl3-SatoEq}. 
Our method is a $q$-analogue of the $3\times 3$-matrix 
formulation of the Painlev\'e VI developed in 
\cite{Har,Maz,Boalch05}.
The technique of the Laplace transform 
has been used to make a connection between 
a $2\times 2$ Fuchsian system and a $3\times 3$
system with irregular singularities \cite{Maz,Boalch05}. 
To construct the $q$-analogue, 
we have used the $q$-Laplace transform 
\eqref{q-Laplace}, which was introduced in \cite{Hahn}.
Note that similar but different versions of 
$q$-Laplace transformations have been 
discussed in several literature \cite{Feinsilver,Zhang}.

We have constructed a class of solutions 
for the $q$-P$_\mathrm{VI}$ 
written in terms of $\bar{W}_0$ (Corollary \ref{cor:solutions}). 
Comparing to the results on the multi-component KP hierarchy 
(see, for example, \cite{UenoTakasaki}),
it may be natural to introduce $\tau$-functions in the 
following manner:
\begin{equation}
(\bar{W}_0(\underline{x}))_{ij}
=\frac{\tau_{ij}(\underline{x})}
{\tau(\underline{x})}\qquad (i,j=1,2,3).
\end{equation}
However, this choice of the $\tau$-functions seems to be 
different form that of \cite{JSRG96,TsudaMasuda}.
It may be important to clarify the relationship 
between the results in \cite{JSRG96,TsudaMasuda} and the 
$q$-$\widehat{\mathfrak{gl}}_3$ hierarchy.

\ack
The authors would like to thank Professors 
K. Hasegawa, G. Kuroki, M. Nishizawa, M. van der Put and H. Sakai
for their interests and discussions. 
The first author is partially supported by 
the Grant-in-Aid for Scientific Research (No.~16740100) from 
the Ministry of Education, Culture, Sports, Science and Technology.
The second author is partially supported by 
the 21st Century COE program of Tohoku University:
Exploring New Science by Bridging Particle-Matter Hierarchy. 

\appendix
\setcounter{section}{1}
\section*{Appendix}
For reader's convenience, we outline a proof of 
the formulas for the $q$-Laplace transformation.
Note that the parameter $q$ is chosen as 
$0<|q|<1$ in \cite{Hahn}, while $|q|>1$ in this paper.
In this appendix, we set $0<|q|<1$ in accordance with \cite{Hahn}, 
and redefine $\mathcal{L}_q$ and $\mathcal{L}^{-1}_q$ as 
\begin{eqnarray}
\mathcal{L}_q[f](\zeta)
 &= \frac{1}{\zeta}
    \sum_{n=0}^\infty
     \frac{q^{n} f(\zeta^{-1}q^{n})}{(q;q)_n},
 \label{q-Laplace_Hahn}\\
\mathcal{L}_q^{-1}[f](z)
 &= \frac{1}{z}
    \sum_{n=0}^\infty
    \frac{(-1)^n q^{n(n-1)/2}f(z^{-1}q^{-n})}{(q;q)_n}.
 \label{Inv_q-Laplace_Hahn}
\end{eqnarray}
If we replace $q$ by $q^{-1}$, 
\eqref{q-Laplace_Hahn} and \eqref{Inv_q-Laplace_Hahn} 
coincide with \eqref{q-Laplace} and \eqref{Inv_q-Laplace},
respectively.

\begin{proposition}
The transformation \eqref{q-Laplace_Hahn} has the 
property, 
\begin{equation}
\label{prop:Lq[Dqf]}
\mathcal{L}_q[\mathcal{D}_{q}(z)f(z)](\zeta)
 = \zeta\mathcal{L}_q[f(z)](\zeta)-(q;q)_\infty^{-1}f(0).
\end{equation}
\end{proposition}
\textit{Proof.} We introduce a truncated version of 
$\mathcal{L}_q$ as 
\begin{equation}
\mathcal{L}^{(M)}_q[f](\zeta)
 = \frac{1}{\zeta}
    \sum_{n=0}^M
     \frac{q^{n} f(\zeta^{-1}q^{n})}{(q;q)_n}.
 \label{q-Laplace_Hahn_finite}
\end{equation}
Then we have 
\begin{eqnarray}
\lefteqn{\mathcal{L}^{(M)}_q[\mathcal{D}_{q}(z)f(z)](\zeta)
 = \sum_{n=0}^M
   \frac{f(\zeta^{-1}q^{n})-f(\zeta^{-1}q^{n+1})}{(q;q)_n}}
\nonumber\\
&= \sum_{n=0}^M
   \frac{f(\zeta^{-1}q^{n})}{(q;q)_n}
 - \sum_{n=0}^M (1-q^{n+1})
   \frac{f(\zeta^{-1}q^{n+1})}{(q;q)_{n+1}}
\nonumber\\
&= \sum_{n=0}^M
   \frac{q^{n}f(\zeta^{-1}q^{n})}{(q;q)_n}
 -(1-q^{M+1})\frac{f(\zeta^{-1}q^{M+1})}{(q;q)_{M+1}}.
\end{eqnarray}
Taking the limit $M\to\infty$, we obtain the formula 
\eqref{prop:Lq[Dqf]}. \qed

The formula \eqref{prop:Lq[Dqf]} coincides with \eqref{formula1} 
by replacing $q$ by $q^{-1}$. The remaining formulas 
\eqref{formula2}--\eqref{formula4} can be obtained in 
similar manner.

\begin{proposition}
\label{prop:inversion}
The transformations \eqref{q-Laplace_Hahn}, 
\eqref{Inv_q-Laplace_Hahn} satisfy 
$\mathcal{L}^{-1}_q[\mathcal{L}_q[f]](z)=f(z)$.
\end{proposition}
\textit{Proof.} 
A straightforward calculation shows that 
\begin{eqnarray}
\mathcal{L}^{-1}_q[\mathcal{L}_q[f]](z)
&= \sum_{i=0}^\infty\sum_{j=0}^\infty
   \frac{(-1)^j q^{i+j(j+1)/2}}{(q;q)_i(q;q)_j}f(xq^{i+j})
\nonumber\\
&= \sum_{k=0}^\infty q^k f(xq^k)
   \sum_{j=0}^k\frac{(-1)^j q^{j(j-1)/2}}{(q;q)_{k-j}(q;q)_j}.
\end{eqnarray}
The result follows from the formula 
\begin{equation}
\label{expansion_q-exp}
(z;q)_k = \sum_{j=0}^k\frac{(q;q)_k}{(q;q)_{j}(q;q)_{k-j}}
(-z)^j q^{j(j-1)/2}, 
\end{equation}
with setting $z=1$. \qed

The relation \eqref{inversion2} can be proved in the same fashion.

\end{document}